\documentclass{article}
\usepackage{amssymb}
\usepackage{amsfonts,amsmath,amscd,graphicx}

\newtheorem{thm}{Theorem}

\newtheorem{cor}{Corollary}
\newtheorem{lem}{Lemma}

\newtheorem{rem}{Remark}

\newcommand{\RR}{{\mathbb R}}

\newcommand{\norm}[1]{\lVert#1\rVert}

\newcommand{\dv}[2]{{\frac{\partial #1}{\partial #2}}}
\newcommand{\dvv}[2]{{\frac{\partial^2 #1}{\partial {#2}^2}}}

\newcommand{\dotex}{{\frac{d}{dt}}}

\begin{document}

\title{A contraction theory-based analysis of the stability of the Extended Kalman Filter}
\author{Silv\`{e}re Bonnabel\thanks{Robotics Center,  Unit\'e Math\'ematiques et Syst\`emes, Ecole des Mines ParisTech, 75272 Paris, France
        ({\tt silvere.bonnabel@mines-paristech.fr}).}
        \and Jean-Jacques Slotine\thanks{Department of Mechanical Engineering, Massachusetts Institute of Technology, MA, 02139 USA ({\tt jjs@mit.edu}).}}

        \maketitle

\begin{abstract}
The contraction properties of the Extended Kalman Filter, viewed as a
deterministic observer for nonlinear systems, are analyzed. This
yields new conditions under which exponential convergence of the state
error can be guaranteed. As contraction analysis studies the evolution
of an infinitesimal discrepancy between neighboring trajectories, and
thus stems from a differential framework, the sufficient convergence
conditions are different from the ones that previously appeared in the
literature, which were derived in a Lyapunov framework. This article
sheds another light on the theoretical properties of this popular
observer.\end{abstract}


\section{Introduction}

Since the seminal work of Kalman and Bucy \cite{Kalman-1961} and
Luenberger \cite{Luenberger}, the problem of building observers for
deterministic linear systems has been laid on firm theoretical
ground. Yet, when the system is nonlinear, there is no general methods
to tackle observer design. Over the last decades, nonlinear observer
design has been an active field of research, and several methods have
emerged for attacking some specific nonlinearities. In the engineering
world, the most popular method is the so-called Extended Kalman Filter
(EKF), a natural extension of the Kalman filter. The principle is to
linearize the system around the trusted (i.e. estimated) trajectory of
the system, build a Kalman filter for this time-varying linear model,
and implement it on the nonlinear model. The EKF is known to yield
good results in practice when the guess on the initial state is close
enough to the actual state, but possesses no guarantee of convergence
in the general case, and indeed can diverge.

Since the 1990's, several papers have addressed the convergence
properties of the EKF viewed as a deterministic observer. Several
conditions under which the estimation error converges to zero have
been derived in, e.g., \cite{baras,song-grizzle-95,boutayeb,Reif}. In
each case, a first set of conditions on the observability and
controllability of the system ensures the boundedness of the solution
of the Riccati equation and of its inverse, and a second set of
conditions ensure in this case the convergence of the estimation error
to zero. Roughly speaking, the latter conditions require either the
initial estimation error to be small, proving the EKF is a local
observer, or the system to be very weakly nonlinear.

In this paper, the convergence properties of the EKF are
studied using contraction theory \cite{slotine-auto} and in particular  the notion of virtual systems \cite{wang-slotine} and virtual observers \cite{jouffroy-slotine}. Historically, ideas closely related
to contraction can be traced back to \cite{lewis49,Hartman,demido} (see e.g. \cite{pde} for a more exhaustive list of references).  In the present case the idea is as follows: instead of studying directly the evolution of the
discrepency, in the sense of a Lyapunov function, between the
estimated state and the true state, contraction theory allows to study the
evolution of the discrepancy between two nearby trajectories of the
EKF, in the sense of a given metric. It is shown that, in a finite
region and under some conditions, two nearby trajectories tend
exponentially towards each other. As a result, the EKF is a dynamical
system which exponentially forgets its initial condition, a very
desirable property for a filter. The fact that the estimation error
tends exponentially to the true state appears then as a mere
consequence of the contraction properties of the filter. Even though
the Lyapunov approach and the contraction approach are based on very
similar metrics, the convergence results obtained in this paper differ
from those of the literature.

The main contributions of this paper are threefold. First, the paper
studies the stability properties of the EKF from the perspective of
contraction theory. This offers an alternative viewpoint to the usual
Lyapunov approach, extending the preliminary results on linear
time-varying systems of~\cite{jouffroy-slotine}. In turn, this
perspective allows simple new convergence results to be derived in
this context (see in particular Theorem 1 and its corollary). Finally,
some of the results are closely related to existing recent literature,
showing both their similarities and their potential strengths.

The paper is organized as follows. In Section \ref{II}, some new general contraction results are derived. Section \ref{sufficient:sec} builds upon those results to derive bounds on the size of the contraction region and the convergence rate of the EKF. Finally, Section \ref{IV} discusses some links with previous work on the stability of the EKF.


\section{General contraction results}\label{II}

Consider the following nonlinear deterministic system
\begin{align}
\dotex  x&=f(x,t)\label{KF1}\\
y_m &=h(x,t)\label{KF2}
\end{align}
where $x\in\RR^n$ is the state, $y_m\in\RR^p$ is the measured output, and $f,h$ are smooth. The EKF equations are given by
\begin{align}
\dotex \hat x&=f(\hat x,t)-PC(\hat x,t)^TR^{-1}(h(\hat x,t)-y_m(t))\label{EKF:eq1}\\
\dotex P&=A(\hat x,t)P(t)+P(t)A(\hat x,t)^T+Q-P(t)C(\hat x,t)^TR^{-1}C(\hat x,t)P(t)\label{EKF:eq2}
\end{align}
where
 $
A( x,t)=\dv{f}{x}( x,t),$ and $C( x,t)=\dv{h}{x}( x,t)
$. In the stochastic theory of Kalman linear filtering, $Q$ and $R$ represent the covariances of the respectively drift noise and measurement noise. In a deterministic and nonlinear setting as the one considered in the present paper, they can be viewed more prosaically as design parameters where $Q^{-1}$ represents the confidence in the trusted model \eqref{KF1} and $R^{-1}$ the confidence in the measurements \eqref{KF2}.  The present analysis relies on the following assumption.

\paragraph*{Assumption 1} From now on we will systematically assume there exist ${\underline{p}}, \overline p>0$ such that  ${\underline{p}} I\leq P(t)\leq{\overline{p}} I$. Moreover, for simplicity we assume that $Q$ is fixed and invertible, and we denote by $\underline q$ its smallest eigenvalue.

The latter assumption on $P(t)$ appears in most papers dealing with the stability of the EKF, e.g.  \cite{baras,deyst,song-grizzle-95}. It is well known that this assumption is verified as soon as the system $\dot \xi  =  A(\hat x(t),t)   \xi   ,   \eta  =  C(\hat x(t),t)   \xi$ is uniformly detectable.
This is of course a very strong prerequisite on the behavior of the filter. Yet, note that  this assumption  can advantageously be checked by the user without any knowledge of the true trajectory. To the authors' best knowledge, very few papers have addressed the stability of the EKF without referring to Assumption 1: see \cite{krener-02} (and more generally high gain observers techniques \cite{deza}) where local convergence results are derived under some different, yet rather restrictive, assumptions.

Let $K(t)=PC(\hat x,t)^TR^{-1}$ denote the Kalman gain.  Consider the
``virtual'' system~\cite{jouffroy-slotine}
\begin{align}\label{virtual:eq}
\dotex z=f(z,t)- K(t)(h(z,t)-y_m(t))
\end{align}
The solution $x(t)$ of the true system \eqref{KF1} is a particular
solution of the virtual system, since for all $t\geq 0$ we have
$h(x(t),t)-y_m(t)=0$. The solution to the EKF equations
\eqref{EKF:eq1}-\eqref{EKF:eq2} is obviously another particular
solution of the virtual system.  As a result, if it can be proven the
distance between two arbitrary trajectories of this
system tends to zero, the convergence of the estimation error $\hat
x-x$ to zero will follow. In turn, this can be achieved by seeking conditions
under which the virtual system~(\ref{virtual:eq}) is contracting.

Let us define a metric for the virtual system~(\ref{virtual:eq}) by choosing, similarly
to the linear time-varying case considered in~\cite{jouffroy-slotine},
the squared length
\begin{align}\label{metric:eq}
\norm{\delta z}^2_{P^{-1}}=\delta z^TP^{-1}\delta z
\end{align}
where $P(t)$ is a solution of the Riccati equation \eqref{EKF:eq2} associated to the trajectory of the estimated state $\hat x(t)$. We have
\begin{align*}
\dotex (\delta z^TP^{-1}\delta z)&=(\dotex \delta z)^TP^{-1}\delta z+\delta z\dot P^{-1}\delta z+\delta zP^{-1}(\dotex\delta z)\\
&=\delta z^T[(A(z,t)-K(t)C(z,t))^TP^{-1}+\dot P^{-1}\\&\qquad+P^{-1}(A(z,t)-K(t)C(z,t))]\delta z
\end{align*}
Using the fact that
 $
\dot P^{-1}=-P^{-1}\dot P P^{-1}
$ where $\dot P$ is given by \eqref{EKF:eq2}, and that
\begin{equation}\begin{aligned}
&[C(z,t)- C(\hat x,t)] R^{-1}[C(z,t)- C(\hat x,t)]\\
&\quad= C^T(z,t) R^{-1} C(z,t)-C^T(\hat x,t) R^{-1} C(z,t)-C^T(z,t) R^{-1} C(\hat x,t)+C^T(\hat x,t) R^{-1} C(\hat x,t)\label{astuce:eq}
\end{aligned}\end{equation}
we finally have
\begin{align}\label{contraction:eq}
\dotex (\delta z^TP^{-1}\delta z)=\delta z^TP^{-1}MP^{-1}\delta z
\end{align}where $
M=P\tilde A^T+\tilde A P +P\tilde C^T R^{-1}\tilde CP-P C^T R^{-1} CP- Q
$,  where we let $\tilde A(z,t)=A(z,t)- A(\hat x,t)~\text{and}~\tilde C(z,t)=C(z,t)- C(\hat x,t)$, and where $C$ denotes the matrix $C(z,t)$.

\subsection{Main result}
Given two symmetric matrices $P_1,P_2$ we define a partial order letting $P_1\leq P_2$ if $P_2-P_1$ is positive semidefinite. We have the following preliminary result:
\begin{lem}Let $0\leq \gamma <{\underline q}/({2\overline p})$. For each time $t\geq 0$ there exists   $r(t)>0$  such that for all $z$ satisfying $ \norm{z-\hat x(t)}\leq r(t)$ we have
\begin{align}\label{MI:eq}
 P\tilde A^T+\tilde A P +P\tilde C^T R^{-1}\tilde CP\leq Q-2\gamma P+P C^T R^{-1} CP
\end{align}
\end{lem}
\paragraph{Proof}
 The inequality is obviously verified for $z=\hat x$ as the right member of \eqref{MI:eq} is a positive definite matrix. As $f,h$ are smooth, the inequality holds in a neighborhood of $\hat x$.

\vspace{.3 cm}

At any time, the vectors $z$ lying within a distance at most $r(t)$ of $\hat x(t)$ are contained in the contraction region, as for those vectors equality \eqref{contraction:eq} becomes the contraction inequality
$$
\dotex (\delta z^TP^{-1}\delta z)\leq-2\gamma(\delta z^TP^{-1}\delta z)
$$This means the squared distance in the sense of metric \eqref{metric:eq} between two neighboring trajectories in this ball will tend to reduce, with a rate of change $\gamma$.  This leads to the following general result:

\begin{thm}\label{thm}Assume there exists $0<\rho=\inf\{r(t),~t\geq 0\}$. Any trajectory of the system \eqref{virtual:eq} that starts in the ball of center $\hat x(0)$ and constant radius $\rho/\sqrt{\overline{p}}$ with respect to the metric \eqref{metric:eq}
 remains in a ball of radius $\rho/\sqrt{\overline{p}}$ centered at the trajectory $\hat x(t)$ of the Extended Kalman Filter \eqref{EKF:eq1}-\eqref{EKF:eq2}, and converges exponentially to this trajectory in the sense of the metric \eqref{metric:eq} with a time constant  $1/\gamma$ for the exponential decay.
 \end{thm}
Mathematically, the theorem's result can be expressed as follows.
Let $d_{P^{-1}}$ be the geodesic distance associated with the metric
\eqref{metric:eq}. Let  $z_1(t),z_2(t)$ be the flows associated to
the system \eqref{virtual:eq} with initial conditions   satisfying
$d_{P^{-1}}(z_i(0),\hat x(0))\leq\rho/\sqrt{\overline p}$ for
$i=1,2$. Then for all times $t\geq 0$
\begin{align}\label{decay:eq}
d_{P^{-1}}(z_1(t),z_2(t))\leq e^{-\frac{t}{\gamma}}d_{P^{-1}}(z_1(0),z_2(0))
\end{align}

\paragraph{Proof}The theorem is a straightforward application of the Theorem 2 of \cite{slotine-auto} which states that any trajectory which starts in a ball of constant radius with respect to the metric centered at a given trajectory and at all times in the contraction region with respect to the metric, remains in that ball and converges exponentially to this trajectory, which is a natural result in the theory of contracting flows (see e.g. \cite{lewis49}). Indeed,  $(z-\hat x)^TP^{-1} (z-\hat x)\leq {\rho^2}/{\overline p}$ implies $\norm{z-\hat x}\leq \rho$ so $z$ is contained in the contraction region  by Lemma 1.

\begin{rem} Note that, if the system is linear, we have $\tilde A(z,t)\equiv\tilde C(z,t)\equiv 0$ for all $z,t$, and we recover the fact that the deterministic Kalman filter for linear systems globally exponentially converges under Assumption 1. \end{rem}

\subsection{Particular case of a linear output map}

The theorem implies an interesting result in the common case of linear
output maps. In many nonlinear systems of engineering interest, the
system output consists of an incomplete measurement of the state
vector. For instance, the output can be temperature or concentrations
in chemical reactors, currents in induction machines, position or
velocity in mechanical systems.  Formally, this means that the output
map is linear, i.e. $h(x)=Cx$, implying $\tilde C(z,t)\equiv 0$ for
all $z,t$. Let then $\lambda_{max}(\cdot)$ denote the largest
eigenvalue of a symmetric matrix, and let $\gamma\geq 0$. The
following result is an immediate consequence of Theorem 1.

\begin{cor}\label{cor1}
Assume that the output map is linear, and  that $\lambda_{max}(\tilde A(z,t)P(t)+P(t)\tilde A(z,t)^T)\leq\underline q-2\gamma \overline p$  for all $z,t$. Then the Extended Kalman Filter \eqref{EKF:eq1}-\eqref{EKF:eq2} is globally exponentially convergent with a time constant $1/\gamma$ for the exponential decay.
\end{cor}

This result shows that contraction analysis can yield new
types of conditions under which exponential convergence of the EKF is
guaranteed. Indeed, under the assumptions of Corollary~\ref{cor1}, the EKF
will converge globally without the standard requirement
that the Hessian of the coordinates of $f$ is uniformly bounded. However, in a more general context, this standard requirement will still be needed as illustrated  in the next section.

\section{A sufficient condition for exponential convergence}\label{sufficient:sec}

We now derive  a lower bound on the size   of the contraction region of Theorem 1. This result relies on usual assumptions on boundedness of second derivatives $\dvv{f}{x},\dvv{h}{x}$ around the observer trajectory $\hat x(t),~t\geq 0$.   We let $\norm\cdot$ and $|\!|\!|\cdot|\!|\!|$ denote the norms on resp. matrices and tensors induced by the Euclidean norm on vectors.

\paragraph*{Assumption 2} There are positive numbers $\alpha,\kappa_A,\kappa_C$ such that for all $z$ satisfying $\norm{\hat x-z}\leq \alpha$ and all $t\geq 0$ we have $|\!|\!|\dvv{f}{x}(z,t)|\!|\!|\leq \kappa_A$ and $|\!|\!|\dvv{h}{x}(z,t)|\!|\!|\leq \kappa_C$.

\vspace{.3 cm}

Under Assumptions 1 and 2, one can derive the following local exponential stability result:

\begin{cor}\label{cor2}Let $\gamma\leq {\underline q}/({2\overline p})$. Let $\zeta^+$ be the positive root of the equation
$$
\frac{\overline p^2}{\underline r} \kappa_C^2\zeta^2+2\overline p \kappa_A\zeta- (\underline q-2\gamma \overline p)=0
$$
Let $\rho=\min(\alpha,\zeta^+)$. Any trajectory of the system
 \eqref{virtual:eq} that starts in the ball of center $\hat x(0)$
  and constant radius $\rho/\sqrt{\overline p}$ with respect to the metric \eqref{metric:eq}
 remains in a ball of radius $\rho/\sqrt{\overline p}$ centered at the trajectory $\hat x(t)$ of the Extended Kalman Filter \eqref{EKF:eq1}-\eqref{EKF:eq2}, and converges exponentially in the sense of the metric \eqref{metric:eq} with a time constant  $1/\gamma$ for the exponential decay.
 In particular, this implies  in the Euclidean metric on vectors, that
\begin{align}\label{eucl:eq}
 \norm{\hat x(t)-x(t)}\leq\sqrt{{\overline p}/{\underline p}}~\norm{\hat x(0)-x(0)}\exp^{-\gamma t}
 \end{align}for all $t\geq 0$ as soon as initially  $\norm{\hat x(0)-x(0)}\leq \rho\sqrt{\underline p}/\sqrt{\overline p}$.
 \end{cor}

\paragraph{Proof}The result directly follows from Theorem \ref{thm}, as long as one can prove that $\rho\leq r(t)$ for all $t\geq 0$, where $r(t)$ is the radius of a ball around $\hat x(t)$ in which  the matrix inequality \eqref{MI:eq} is verified. To begin, note that the ball of center $\hat x$ and radius $\zeta^+$ is equivalently defined as the set
\begin{align}\label{set}
\{z\in\RR^n,~  2\overline p \kappa_A\norm{\hat x-z}+\frac{\overline p^2}{\underline r} \kappa_C^2\norm{\hat x-z}^2\leq \underline q-2\gamma\overline p\}
\end{align}
Let $\tilde e=z-\hat x$. As
 $
\tilde A(z,t)=A(z,t)- A(\hat x,t)=\int_0^1\dvv{f}{x}(\hat x+r\tilde e,t)\tilde e ~dr
 $
and
 $
\tilde C(z,t)=C(z,t)- C(\hat x,t)=\int_0^1\dvv{h}{x}(\hat x+r\tilde e,t)\tilde e ~dr
$,
we have $\norm{\tilde A}\leq \kappa_A\norm{\hat x-z}$ and $\norm{\tilde C}\leq \kappa_C\norm{\hat x-z}$. The largest value of the symmetric matrix $\tilde AP+P\tilde A^T$ satisfies
\begin{align*}
\lambda_{max}(\tilde AP+P\tilde A^T)&=\max_{\norm w=1} w^T(\tilde AP+P\tilde A^T)w
\\&=2\max_{\norm w=1}\text{Trace}(w^T\tilde APw)
\\&\leq 2\overline p \max_{\norm w=1}w^T\tilde Aw
\\&\leq 2\overline p  \max_{\norm w=1,\norm v=1}|w^T\tilde Av|\leq 2\overline p\norm {\tilde A}\leq 2\overline p \kappa_A\norm{\hat x-z}
\end{align*}
In the same way we have $ \lambda_{max}(P\tilde C^T R^{-1}\tilde
CP)\leq \frac{\overline p^2}{\underline r} \kappa_C^2\norm{\hat
x-z}^2 $. Thus,  as long as $z$ belongs to the set  \eqref{set}, the
inequality \eqref{MI:eq} is satisfied.

\section{Links with previous work in the literature}\label{IV}

The Extended Kalman filter has been shown to converge locally
exponentially under a set of conditions on the nonlinearities of the
system, see e.g. \cite{Reif,deyst} in continuous-time and
\cite{song-grizzle-95,boutayeb,baras} in discrete-time. In these
papers, the convergence analysis is based on the Lyapunov function
\begin{align}\label{lyapunov:eq}
V(x-\hat x,t)=(x-\hat x)^TP^{-1}(t)(x-\hat x)
\end{align}
At fixed time $t$, $V(x-\hat x,t)$ is the geodesic distance between $x$ and $\hat x$ for the proposed metric
$$\norm{\delta z}^2_{P^{-1}}=\delta z^TP^{-1}\delta z
$$ Thus, it is no surprise that the convergence conditions derived in this
article are very similar to those previously appearing in the
literature. That said, the specificity of the contraction
framework yields some differences that we shall detail in this
section, which is organized as follows. In Subsection \ref{A}, we
emphasize a difference of point of view between contraction analysis
and Lyapunov based convergence analysis. In Subsection \ref{B}, we compare
the convergence rate and basin of attraction of Section
\ref{sufficient:sec} with the results appearing previously in the
literature. For fair comparison we chose the article \cite{Reif}
which deals with the continuous-time case. Finally, in Subsection \ref{C},
the modification of the EKF proposed in \cite{Reif} is
analyzed in the light of contraction theory, yielding a simple
generalization of this work which is straightforward in our framework.

\subsection{A different approach}\label{A}

In standard Lyapunov analysis of nonlinear deterministic observers,
one generally seeks to prove the state error, i.e., the discrepancy
between the estimated trajectory $\hat x$ and the true trajectory $x$,
measured by some Lyapunov function, tends to zero. Often the observers
can be only proved to be locally convergent, i.e., the initial guess
$\hat x(0)$ must belong to some attraction basin containing $x(0)$.
The contraction analysis of the present paper, based on the idea of
\emph{virtual systems}~\cite{wang-slotine} and specifically \emph{virtual
observers}~\cite{jouffroy-slotine}, builds upon a different
approach. Indeed, the idea is to focus on a particular trajectory
 of the filter $\hat x(t),~t\geq 0$, and then to study the evolution of
two neighboring trajectories of the virtual system \eqref{virtual:eq},
and prove that the distance between them tends to shrink over the
time. Under a set of conditions, we have proved that this property holds in
a ball centered at the particular trajectory $\hat x(t),~t\geq 0$.

The contraction results underlying Theorem \ref{thm} and its
corollaries are natural properties to expect from any filter. Indeed,
they prove that as a dynamical system the observer possesses
stability properties, and this independently from the behavior of the
true system, or modeling errors. To be more specific, it is proved two initially close enough trajectories of the filter equation will verify equation \eqref{decay:eq}. This
exponential forgetting of the initial condition is thus a very
desirable property for filters, which indicates robustness with respect
to bad guesses and perturbations.  This property is valid whether the
true trajectory $x(t)$ belongs to the contraction set or not.

Besides, contraction theory provides a  concrete result on  the robustness of the EKF against external perturbations. Suppose indeed that  $x_p(t)$ is some trajectory of a \emph{perturbed} virtual system $\dotex x_p=f(x_p,t) - K(t)(h(x_p,t)-y_m(t))+b(x_p,t)$ where $b(x_p,t)$ represents  a disturbance whose norm is supposed to be uniformly bounded by, say, $\norm{b}_{\text{max}}$.  Then we have (see  \cite{slotine-auto,delvecchio-slotine}): $\norm{x_p(t)-\hat x(t)}\leq \sqrt{(\overline p/\underline p)}\bigl( e^{-\frac{t}{\gamma}}\norm{x_p(0)-\hat x(0)}+ \gamma \norm{b}_{\text{max}}\bigr)$.  It proves that any trajectory of the perturbed system converges exponentially to a ball of radius $\sqrt{(\overline p/\underline p)}\gamma \norm{b}_{\text{max}}$ around the observer trajectory, allowing to evaluate the estimation error generated by the perturbation. The latter result completes the robustness properties of the EKF.

\subsection{Comparison of convergence results}\label{B}

First of all, to the authors' knowledge Theorem 1 and its Corollary 1
have never appeared in the literature.  The common approach
is to study the evolution of the Lyapunov function
\eqref{lyapunov:eq} over time. While similar results to Theorem 1 and
its corollary may possibly be worked out from this approach as well,
they appear quite naturally in a contraction framework.

Consider now the results of Section \ref{sufficient:sec}. As already
mentioned, the standard Lyapunov function \eqref{lyapunov:eq}
represents the geodesic distance between $\hat x$ and $x$ in the sense
of the metric proposed in this paper.  This is why the bounds derived
in Section \ref{II} are very similar to those previously obtained in the
literature. However, they are different, and this is mainly due to the
use of equation \eqref{astuce:eq} in the result
\eqref{contraction:eq}. Again, the analog of this transformation in the
Lyapunov framework is not easily seen, whereas it appears naturally in
the contraction framework.

Assume for simplicity that $\alpha=+\infty$. Table \ref{comparison:table} compares the convergence rate and attraction basin obtained in \cite{Reif} and the ones obtained in the present paper in the two limiting cases $\kappa_A=0$ and $\kappa_C=0$. The results all correspond to the error equation \eqref{eucl:eq} in the Euclidean metric. The rates and bounds in the Lyapunov approach of \cite{Reif} have been obtained letting $\alpha=0$ (this parameter has a different meaning in this article) and using the definition of $\kappa$ derived therein.

\begin{table}
\center{
\begin{tabular}{|l|c|r|r|}
  \hline
   & \textbf{Convergence rate} & \textbf{Attraction basin for $\kappa_C=0$} &\textbf{Attraction basin for $\kappa_A=0$} \\
  \hline Lyapunov  & $\gamma=\underline q\underline p/(4\overline p^2)$ & $\norm{x(0)-\hat x(0)}\leq (\underline p /\overline p) ~[\underline q  /(4 \kappa_A\overline p) ]$ &$\norm{x(0)-\hat x(0)}\leq \underline q \underline r /(4\overline c\kappa_C\overline p^2)$\\
  \hline
Contraction &$\gamma=\underline q/(4\overline p)$ &$\norm{x(0)-\hat x(0)}\leq (\underline p /\overline p)^{1/2} [ \underline q  /(4 \kappa_A\overline p)]$ &$\norm{x(0)-\hat x(0)}\leq (\underline q{\underline p}\underline r)^{1/2}  /( \kappa_C\overline p^{3/2}\sqrt 2)$ \\
    \hline
\end{tabular}}\caption{Comparison of convergence rate and attraction basin. }\label{comparison:table}
\end{table}

We see that the results are quite similar, yet different.  First of
all, we immediatly see that both the  convergence rate and the size of the guaranteed attraction basin are larger in our approach in the limiting case $\kappa_C=0$, improving the formerly obtained bounds of  \cite{Reif}. Indeed $\gamma$ given in
Table \ref{comparison:table} is greater in the contraction case by a factor $\overline
p /\underline p\geq 1$ (yielding faster convergence), and the size of the attraction basin  by a factor $(\overline
p /\underline p)^{1/2}\geq 1$.  The difference is much more remarkable
in the limit case $\kappa_A=0$, in which case the size of the
attraction basin depends heavily on the problem parameters. In
particular, a noticeable difference is that the size of the attraction
basin in \cite{Reif} depends on $\overline c$, an upper bound on
$\{\norm{C(t)},~t\geq 0\}$. As a result, a large linearized
observation matrix $C(t)$ can diminish the guaranteed size of the
attraction basin. On the other hand, the
bounds obtained in the present paper do not rely on an upper bound
$\overline c$, and therefore they may yield stronger guarantees in some cases.

 \subsection{A contraction-based interpretation  of the observer of \cite{Reif} }\label{C}

In \cite{Reif}, the authors propose to modify slightly the EKF  by adding a new term $2\beta P$ in the Riccati equation  \eqref{EKF:eq2},  with $\beta\geq 0$, leading to a modified Riccati equation:
$$
\dotex P=A(\hat x,t)P(t)+PA(\hat x,t)^T+Q-PC(\hat x,t)^TR^{-1}C(\hat x,t)P +2\beta P
$$
They prove the addition of such a term yields a faster convergence rate, as long as Assumption 1 is preserved. Note that, such a term tends to increase the eigenvalues of $P$ and thus to destabilize the Riccati equation. The fact that Assumption 1 remains then  valid is thus non trivial and must be checked. Then, as long as Assumption 1 is proved to hold,  this term ensures a faster convergence rate.
This latter fact is easily understood in the contraction framework. Indeed, with this additional term, inequality \eqref{MI:eq} becomes:
$$
P\tilde A^T+\tilde A P +P\tilde C^T R^{-1}\tilde CP\leq Q-2(\gamma+\beta) P+P C^T R^{-1} CP$$
The benefits of this term are now obvious as it transforms the convergence rate $\gamma$ into $\gamma+\beta$. In fact,  contraction theory even offers a direct generalization of the work of \cite{Reif}. Indeed consider the following modified EKF observer
\begin{align*}
\dotex \hat x&=f(\hat x,t)-PC(\hat x,t)^TR^{-1}(h(\hat x,t)-y_m(t))\\
\dotex P&=A(\hat x,t)P(t)+P(t)A(\hat x,t)^T+Q-P(t)C(\hat x,t)^TR^{-1}C(\hat x,t)P(t)+2N
\end{align*}
where $N$ is any positive definite matrix. It follows directly from equation \eqref{MI:eq} that as long as Assumption 1 still holds, the guaranteed convergence rate of this observer is increased by adding  $\underline n /\overline p$, where $\underline n$ denotes the smallest eigenvalue of $N$.

\subsection*{Acknowledgements}The authors would like to thank Laurent Praly for interesting feedback on the paper.
\bibliographystyle{plain}

\end{document}